\documentclass[aps,prd,preprint,tightenlines,superscriptaddress]{revtex4}
\usepackage{epsfig}
\usepackage{graphicx}
\usepackage{psfrag}
\usepackage{amsmath,amssymb}
\usepackage{colordvi}
\usepackage{color}
\usepackage{amsfonts}
\usepackage{enumerate}
\usepackage{slashed,bm}

\def \non {\nonumber}
\def \beq  {\begin{equation}}
\def \eeq  {\end{equation}}

\begin{document}

\title{One-loop matching for transversity\\
generalized parton distribution}

\author{Xiaonu Xiong}

\affiliation{Istituto Nazionale di Fisica Nucleare, Sezione di Pavia, Pavia, 27100,
Italy}

\author{Jian-Hui Zhang}


\affiliation{Institut f\"ur Theoretische Physik, Universit\"at Regensburg, \\
 D-93040 Regensburg, Germany}
\begin{abstract}
Recent developments showed that light cone parton distributions can be studied by investigating the large momentum limit of the so-called quasiparton distributions, which are defined in terms of spacelike correlators, and therefore can be readily computed on the lattice. These two distributions can be connected to each other by a perturbative factorization formula or matching condition that allows one to convert the latter into the former. Here we present the one-loop matching condition for the transversity generalized quark distribution in the nonsinglet case. 
\end{abstract}
\maketitle

\section{introduction}

Understanding the internal structure of the proton is an important goal of hadron physics. Although the fundamental constituents of the proton--quarks and gluons--can be well described by the QCD Lagrangian, we are still lacking a systematic framework enabling us to fully calculate the proton properties from its quark and gluon constituents. Therefore, we have to resort to phenomenological functions to characterize the proton structure and determine them by fitting to experimental data. One example of such functions is the parton distribution functions, which characterize the momentum distribution of quarks and gluons inside the proton, and play a crucial role in computing physical cross sections at hadron colliders such as the Large Hadron Collider.

The parton distributions are defined as the forward hadronic matrix element of nonlocal light cone correlations where the initial and final hadron have identical four-momenta. They have been intensively studied in the literature in the past few decades. In recent years, their generelization to nonforward kinematics, the generalized parton distributions (GPDs)~\cite{Ji:1996ek, Radyushkin:1996nd, Ji:1996nm, Radyushkin:1996ru}, also received considerable attention. In contrast to the parton distributions, the
GPDs encode more information about the internal structure of nucleons, and can shed light on the three-dimensional spatial picture~\cite{Burkardt:2000za,Diehl:2002he,Ralston:2001xs} and the spin structure of the nucleon~\cite{Ji:1996ek}. Experimentally, the GPDs can be accessed in exclusive processes such as deeply virtual Compton scattering or meson production. However, they are rather difficult to access theoretically from lattice QCD, since their definition explicitly involves light cone correlations.


Recent developments~\cite{Ji:2013fga, Ji:2013dva, Hatta:2013gta, Xiong:2013bka, Lin:2014zya, Ma:2014jla, Ji:2014gla, Ji:2014hxa, Ji:2014lra, Ji:2015jwa, Alexandrou:2015rja, Ji:2015qla} showed that hadronic matrix elements involving light cone correlations can be studied by moving the hadron slightly off the light cone and then boosting back. In the case of parton distributions, their light cone definition can be approached by first considering the hadronic matrix element of suitable spacelike correlations at a finite but large hadron momentum and then taking the infinite momentum limit. Of course, taking the infinite momentum limit does not directly yield the light cone result, as it contains a singular dependence on the hadron momentum. However, this singular momentum dependence can be traded into the renormalization scale dependence of the light cone result by a perturbative factorization formula or matching condition. Some explicit examples of such a matching have been given in Refs.~\cite{Ji:2013fga, Xiong:2013bka}, where the perturbative matching factor was computed up to one-loop level. The advantage of the above approach is that it offers a practical possibility to tackle the difficult task of computing hadronic matrix elements of light cone correlations by dividing it into two parts that are separately computable: the matrix elements of spacelike correlations at a finite hadron momentum can be computed on the lattice, and the matching condition is perturbative.


In Ref.~\cite{Ji:2015qla}, we have considered the one-loop matching for the unpolarized and longitudinally polarized GPDs. The main purpose of the present paper is to establish the one-loop matching
for the quark's transversity GPD, which is defined through the
following parametrization of the nonforward nucleon matrix element~\cite{Diehl:2001pm}
\begin{align}
F_{q}^{T}\left(x,\xi,t\right)= & \int\frac{dz^{-}}{4\pi}e^{ixp^{+}z^{-}}\langle p''|\bar{\psi}(-\frac{z}{2})i\sigma^{+\perp}L\left(-\frac{z}{2},\frac{z}{2}\right)\psi(\frac{z}{2})|p'\rangle_{z^{+}=0,\vec{z}_{\perp}=0}\nonumber \\
= & \frac{1}{2p^{+}}\left[H_{T}(x,\xi,t)\bar{u}(p'')i\sigma^{+\perp}u(p')+\tilde{H}_{T}(x,\xi,t)\bar{u}(p'')\frac{p^{+}\Delta^{\perp}-\Delta^{+}p^{\perp}}{m^{2}}u(p')\right.\nonumber \\
 & \left.+E_{T}(x,\xi,t)\frac{\gamma^{+}\Delta^{\perp}-\Delta^{+}\gamma^{\perp}}{2m}u(p')+\tilde{E}_{T}(x,\xi,t)\bar{u}(p'')\frac{\gamma^{+}p^{\perp}-p^{+}\gamma^{\perp}}{m}u(p')\right],
\end{align}
where $L\left(-\frac{z}{2},\frac{z}{2}\right)$ is the gauge link
along the light cone and
\begin{align}
p^{\mu}= & \frac{p''^{\mu}+p'^{\mu}}{2},\;\;\Delta^{\mu}=p''^{\mu}-p'^{\mu},\;\;\xi=\frac{p''^{+}-p^{'+}}{p''^{+}+p^{'+}}, \;\; t=\Delta^2.
\end{align}
In the forward limit $\xi, t\to 0$, $H_{T}\left(x,\xi,t\right)$
reduces to the quark transversity distribution $\delta q\left(x\right)$, while
$\tilde{H}_{T}(x, \xi ,t)$ and $E_{T}(x, \xi, t)$ are absent because they are associated with the momentum
transfer $\Delta^\mu$. Also $\tilde E_T(x, \xi, t)$ drops out due to the Gordon identity. Unlike the unpolarized and longitudinally polarized
GPD that preserve quark helicity, the transversity GPD describes quark helicity flip
due to the $\sigma^{+\perp}$ structure
in the bilocal field correlator which also appears in the quark transversity
distribution. Therefore, the transversity GPD is a chiral-odd distribution and difficult to probe in experiments, since to access it requires hard processes allowing for the quark to change its chirality, e.g., the
double vector meson photoproduction$\gamma_{T}^{*}N\rightarrow\rho_{L}\rho_{T}N'$
and the exclusive $\pi^{0},\eta$ electroproduction $\gamma^{*}N\rightarrow\pi^{0}(\eta)N'$ ~\cite{Ivanov:2002jj,Enberg:2006he,Goldstein:2013vda,Pire:2015iza}.
The transversity GPDs $\tilde{H}$ and $E$ can be related to the quark
contribution to the nucleon transverse anomalous magnetic moment by~\cite{Burkardt:2005hp}
\begin{align}
\kappa_{T}^{q}= & \int_{\xi-1}^{1}dx\,\left[2\tilde{H}_{T}^{q}\left(x,\xi,t\right)+E_{T}^{q}\left(x,\xi,t\right)\right]_{\xi=0,t=0}.
\end{align}
The impact parameter space probability interpretation of GPDs can also
be extended to the chiral-odd GPDs: the two-dimensional Fourier transform
of the combination $2\tilde{H}_{T}^{q}(x,\xi=0,-\vec{\Delta}_{\perp}^{2})+E_{T}^{q}(x,\xi=0,-\vec{\Delta}_{\perp}^{2})$
has been shown to measure the distortion of quark distribution on the transverse impact
parameter plane inside a transversely polarized nucleon~\cite{Burkardt:2005hp}.

The rest of the paper is organized as follows. In Section II, we present the definition of the quasitransversity
GPD, and the one-loop results for the quasi and light cone transversity GPD in the nonsinglet case. In Section III, we present the factorization formula for the quasitransversity GPD and the one-loop matching factor. Section IV contains our conclusion.

\section{one-loop result for transversity quark gpds}
According to Ref.~\cite{Ji:2013dva}, the quasiquark transversity GPD can be defined in complete analogy with its light cone counterpart as
\begin{align}\label{QuaDef}
\mathcal{F}_{q}^{T}\left(x,\xi,t\right)= & \int\frac{dz}{4\pi}e^{-ik^{z}z}\langle p''|\bar{\psi}(-\frac{z}{2})i\sigma^{z\perp}L\left(-\frac{z}{2},\frac{z}{2}\right)\psi(\frac{z}{2})|p'\rangle_{z^{0}=0,\vec{z}_{\perp}=0}\nonumber \\
= & \frac{1}{2p^{z}}\left[\mathcal{H}_{T}(x,\xi,t,p^{z})\bar{u}(p'')i\sigma^{z\perp}u(p')+\tilde{\mathcal{H}}_{T}(x,\xi,t,p^{z})\bar{u}(p'')\frac{p^{z}\Delta^{\perp}-\Delta^{z}p^{\perp}}{m^{2}}u(p')\right.\nonumber \\
 & \left.\mathcal{E}_{T}(x,\xi,t,p^{z})\frac{\gamma^{z}\Delta^{\perp}-\Delta^{z}\gamma^{\perp}}{2m}u(p')+\tilde{\mathcal{E}}_{T}(x,\xi,t,p^{z})\bar{u}(p'')\frac{\gamma^{z}p^{\perp}-p^{z}\gamma^{\perp}}{m}u(p')\right].
\end{align}
The gauge link $L$ now points along the spatial $z$ direction. We denote
\begin{equation}
p'^{\mu}=p^{\mu}-\frac{\Delta^{\mu}}{2},\;\;\;\;p''^{\mu}=p^{\mu}+\frac{\Delta^{\mu}}{2},\;\;\;\;p^{\mu}=\left(p^{0},0,0,p^{z}\right),\;\;\;\;\xi=\frac{p''^{z}-p'^{z}}{2p^{z}}=\frac{\Delta^{z}}{2p^{z}},
\end{equation}
$t=\Delta^{2}$ is the same as in the light cone GPD since it is Lorentz
invariant, and the skewness parameter $\xi$ is now defined in terms of the $z$ component of the external momenta. By construction, the quasi GPD defined above approaches the light cone one in the limit $p^z\to\infty$. However, in a practical computation of hadronic matrix elements (e.g. in lattice computations), one has to start with a finite $p^z$. The result then exhibits a different UV behavior from that of the light cone one, because the limit $p^z\to\infty$ and UV regularization are not interchangeable. However, these two results can be related to each other by a perturbative factorization formula or matching condition up to corrections suppressed by powers of $p^z$~\cite{Ji:2013dva}. Note that in the quasi distribution, the functions $\mathcal{H}_{T}, \mathcal{\tilde{H}}_{T}, \mathcal{E}_{T}$ and $\tilde{\mathcal{E}}_{T}$ may have $p^{z}$ dependence at a finite $p^z$. In order not to lose generality, we choose $\vec{\Delta}_{\perp}$
to have both $x$- and $y$ components, as the nucleon is now transversely polarized. 
$\xi$ is constrained by the requirement $\vec{\Delta}_{\perp}^2\geq0$ which leads to
\begin{align}
0<\xi<\frac{1}{2p^{z}}\sqrt{\frac{-t\left(\left(p^{z}\right)^{2}+m^{2}-\frac{t}{4}\right)}{m^{2}-\frac{t}{4}}}\overset{{\scriptstyle p^{z}\rightarrow\infty}}{\rightarrow} & \sqrt{\frac{-t}{-t+4m^{2}}}\ .
\end{align}
Here we assume $\xi>0$; the case $\xi<0$ can be related to $\xi>0$ by time reversal invariance.

Let us first take a quark as the external state. It is straightforward to show that the quasi- and light cone transversity GPD yield the same result at tree level, where
\begin{align}
H_{T}^{\left(0\right)}\left(x,\xi,t\right)={\cal H}_{T}^{\left(0\right)}\left(x,\xi,t,p^{z}\right)= & \delta\left(1-x\right),
\end{align}
while all the other functions $\tilde{H}_{T}$, $\tilde{\mathcal{H}}_{T}$,
$E_{T}$, $\mathcal{E}_{T}$, $\tilde E_{T}$, and $\mathcal{\tilde{E}}_{T}$ vanish
at this level.

\begin{figure}
\centering \includegraphics[scale=0.4]{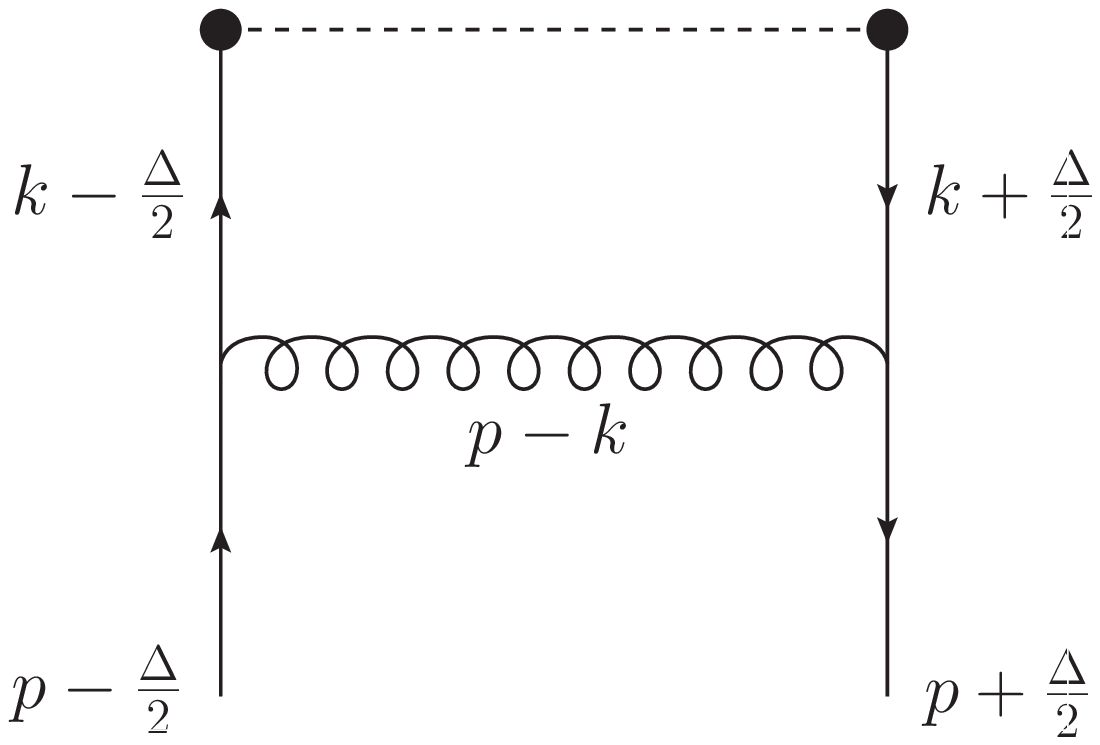} \hspace{0.2em}
\includegraphics[scale=0.6]{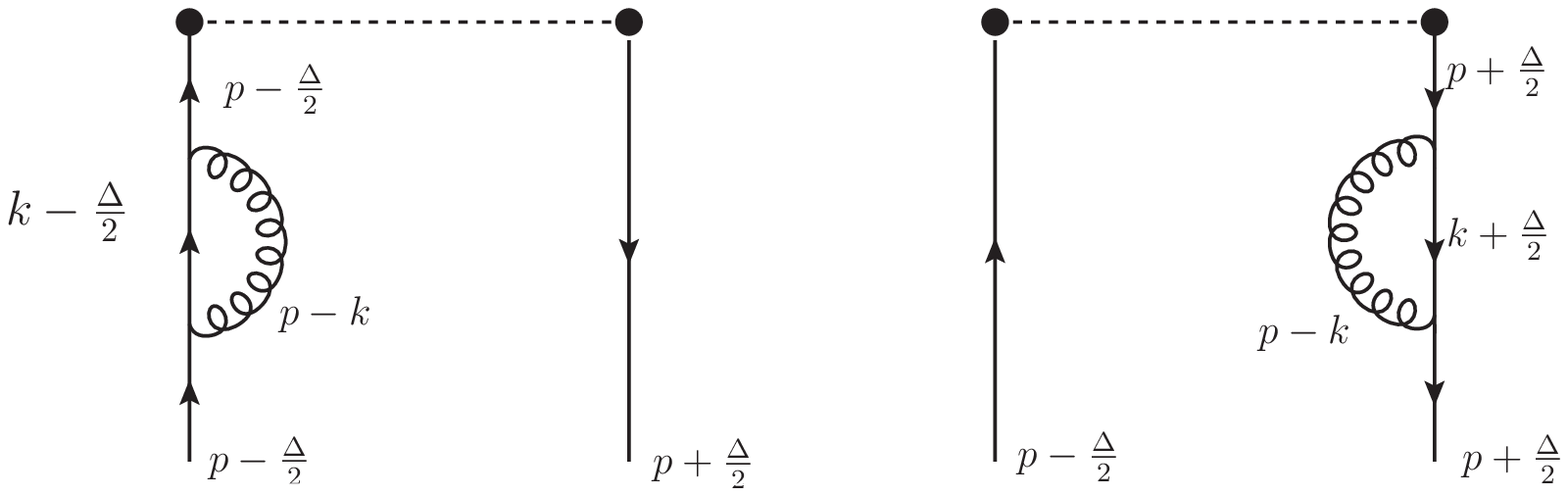} \protect\caption{One-loop diagrams for quark transversity GPD.}
\label{1loopGPD}
\end{figure}

The one-loop calculation can in principle be carried out in any gauge, since the definition is gauge invariant. We will choose the axial gauge $A^z=n\cdot A=0$, in which the gauge link becomes unity. In this gauge, the contributing Feynman diagrams are shown in Fig.~\ref{1loopGPD}, and the gluon propagator is
\begin{equation}\label{gluonprop}
\frac{-i}{k^{2}+i\epsilon}\left[g_{\mu\nu}-\frac{n_{\mu}k_{\nu}+n_{\nu}k_{\mu}}{n\cdot k}+n^{2}\frac{k_{\mu}k_{\nu}}{\left(n\cdot k\right)^{2}}\right],
\end{equation}
where $n\cdot k=k^{z}$ and $n^{2}=-1$. If one chooses to work
in the covariant Feynman gauge, the second and last term in the above gluon propagator will correspond to the gauge link diagrams in the Feynman gauge and will yield the same one-loop result.

In the following, we separately compute the contribution coming from the three terms in the gluon propagator in Eq.~(\ref{gluonprop}) and regularize the UV divergences by a transverse momentum cutoff $\mu$. The $g^{\mu\nu}$ term leads to
\begin{align}
\notag \Gamma_{1}= & C_{F}\int\frac{d^{4}k}{\left(2\pi\right)^{4}}\bar{u}\left(p+\frac{\Delta}{2}\right)\left(-ig_s\gamma^{\mu}\right)\frac{i}{k\!\!\!/+\frac{\Delta\!\!\!/}{2}-m}i\sigma^{z1}\frac{i}{k\!\!\!/-\frac{\Delta\!\!\!/}{2}-m}\left(-ig_s\gamma^{\nu}\right)u\left(p-\frac{\Delta}{2}\right)\\
\notag & \times\frac{-ig_{\mu\nu}}{\left(p-k\right)^{2}}\delta\left(x-\frac{k^{z}}{p^{z}}\right)\\
\notag = & iC_{F}g_{s}^{2}\int\frac{d^{4}k}{\left(2\pi\right)^{4}}\frac{1}{\left[\left(k+\frac{\Delta}{2}\right)^{2}-m^{2}\right]\left[\left(k-\frac{\Delta}{2}\right)^{2}-m^{2}\right]\left(p-k\right)^{2}}\bar{u}\left(p+\frac{\Delta}{2}\right)\\
\notag & \times\left\{ -\left(\gamma^{\perp}\gamma^{z}k\!\!\!/\Delta\!\!\!/+\Delta\!\!\!/k\!\!\!/\gamma^{z}\gamma^{\perp}\right)+\left(\gamma^{\perp}\gamma^{z}\Delta\!\!\!/k\!\!\!/+k\!\!\!/\Delta\!\!\!/\gamma^{z}\gamma^{\perp}\right)\right.\\
&\left.-m\left[\left(\gamma^{\perp}\gamma^{z}\Delta\!\!\!/-\Delta\!\!\!/\gamma^{\perp}\gamma^{z}\right)+2\left(\gamma^{\perp}\gamma^{z}k\!\!\!/+k\!\!\!/\gamma^{\perp}\gamma^{z}\right)\right]\right\} u\left(p-\frac{\Delta}{2}\right)\delta\left(x-\frac{k^{z}}{p^{z}}\right).
\end{align}
After introducing a Feynman parametrization and integrating over $k$, one has
\begin{align}
\notag \Gamma_{1}= & -\frac{C_{F}g_{s}^{2}}{8\pi^2}\int_{0}^{1}\!\!d\lambda\int_{0}^{1-\lambda}\!\!\!d\eta\frac{1}{2\left[\left(p_{z}\right)^{2}\left(1\!-\!x\!-\!\lambda\left(1\!+\!\xi\right)\!-\!\eta\left(1-\xi\right)\right)^{2}\!+\!\!m^{2}\left(\lambda\!+\!\eta\right)^{2}\!-\!t\lambda\eta\right]^{3/2}}\\
\notag & \times\bar{u}\left(p+\frac{\Delta}{2}\right)\left\{ \left(-t+2m^{2}\right)\left(1-\lambda-\eta\right)i\sigma^{z\perp}+2m^{2}\left[x-\left(1-\lambda-\eta\right)-\left(\lambda-\eta\right)\xi\right]\frac{p^{z}\Delta^{\perp}}{m^{2}}\right.\\
&\left.-2m^{2}\left(\lambda+\eta\right)\frac{\gamma^{z}\Delta^{\perp}-\Delta^{z}\gamma^{\perp}}{2m}-2m^{2}\left(\lambda-\eta\right)\frac{-p^{z}\gamma^{\perp}}{m}\right\} u\left(p-\frac{\Delta}{2}\right).
\end{align}
The spinor structure has been rewritten as the combination of $\mathcal{H}$,
$\tilde{\mathcal{H}}$, $\mathcal{E}$ and $\tilde{\mathcal{E}}$
in Eq.(\ref{QuaDef}) using the Gordon identity and Dirac equation. Integrating
out the Feynman parameters gives
\begin{align}
\mathcal{H}_{T,1}\left(x,\xi,t,p^{z}\right)&=\frac{\alpha_{s}C_F}{2\pi} \begin{cases}
\left(\frac{x}{1-x}+\frac{\xi}{1+\xi}\right)\ln\frac{m^{2}}{-t}+\left(\frac{x}{1-x}-\frac{\xi^{2}}{1-\xi^{2}}\right)\ln\frac{\xi-x}{\xi+x}\\
-\frac{\xi}{1-\xi^{2}}\ln\frac{\left(1+\xi\right)^{2}\left(\xi^{2}-x^{2}\right)}{4\left(1-x\right)^{2}\xi^{2}} & -\xi<x<\xi\\
\frac{2\left(x-\xi^{2}\right)}{\left(1-x\right)\left(1-\xi^{2}\right)}\ln\frac{m^{2}}{-t}+\frac{2\xi}{1-\xi^{2}}\ln\frac{1-\xi}{1+\xi} & \xi<x<1\\
0 & \text{Otherwise},
\end{cases}\non\\
%
\tilde{\mathcal{H}}_{T,1}\left(x,\xi,t,p^{z}\right)&=\mathcal{O}\left(\frac{m^2}{\left(p^z\right)^2}\right),\non\\
\mathcal{E}_{T,1}\left(x,\xi,t,p^{z}\right)&= \frac{\alpha_{s}C_{F}}{2\pi}\frac{2m^{2}}{-t}\begin{cases}
\frac{1}{1+\xi}\ln\frac{-t}{m^{2}}-\frac{2\xi}{1-\xi^{2}}\ln\frac{1+\xi}{2\xi(1-x)}\\+\frac{1}{1+\xi}\ln(x+\xi)-\frac{1}{1-\xi}\ln(\xi-x) & -\xi<x<\xi\\
\frac{2}{1-\xi^{2}}\ln\frac{-t}{m^{2}}-\frac{2\xi}{1-\xi^{2}}\ln\frac{1+\xi}{1-\xi} & \xi<x<1\\
0 & \text{otherwise},
\end{cases}\non\\
\mathcal{\tilde{E}}_{T,1}\left(x,\xi,t,p^{z}\right)&=	\frac{\alpha_{s}C_F}{2\pi}\frac{m^{2}}{-t}\begin{cases}
-\frac{2}{1+\xi}\ln\frac{-t}{m^{2}}-\frac{2\xi}{1-\xi^2}\ln\frac{\xi-x}{\xi+x}\\
-\frac{2}{1-\xi^{2}}\ln\frac{\left(1+\xi\right)^{2}\left(\xi^{2}-x^{2}\right)}{4\left(1-x\right)^{2}\xi^{2}} & -\xi<x<\xi\\
\frac{4}{1-\xi^{2}}\left(\xi\ln\frac{-t}{m^{2}}+\ln\frac{1-\xi}{1+\xi}\right) & \xi<x<1\\
0 & \text{otherwise}.
\end{cases}
\end{align}
From the above results, $\tilde{\mathcal{H}}_{T,1}$ is power suppressed by $p^z$ and will be omitted.

The second term in the gluon propagator gives
\begin{align}
\notag \Gamma_{2}=	& C_{F}\int\frac{d^{4}k}{\left(2\pi\right)^{4}}\bar{u}\left(p+\frac{\Delta}{2}\right)\left(-ig_{s}\gamma^{\mu}\right)\frac{i}{k\!\!\!/+\frac{\Delta\!\!\!/}{2}-m}i\sigma^{z\perp}\frac{i}{k\!\!\!/-\frac{\Delta\!\!\!/}{2}-m}\left(-ig_{s}\gamma^{\nu}\right)u\left(p-\frac{\Delta}{2}\right)\\
\notag
 &\times\frac{i\left[\left(p-k\right)^{\mu}n^{\nu}+\left(p-k\right)^{\nu}n^{\mu}\right]}{n\cdot\left(p-k\right)\left(p-k\right)^{2}}\delta\left(x-\frac{k^{z}}{p^{z}}\right)\\
\notag =&	iC_{F}g_{s}^{2}\int\frac{d^{4}k}{\left(2\pi\right)^{4}}\bar{u}\left(p+\frac{\Delta}{2}\right)\left[\gamma^{z}\gamma^{\perp}\frac{1}{k\!\!\!/-\frac{\Delta\!\!\!/}{2}-m}\gamma^{z}+\gamma^{z}\frac{1}{k\!\!\!/+\frac{\Delta\!\!\!/}{2}-m}\gamma^{z}\gamma^{\perp}\right]u\left(p-\frac{\Delta}{2}\right)\\
&\times\frac{1}{n\cdot\left(p-k\right)\left(p-k\right)^{2}}\delta\left(x-\frac{k^{z}}{p^{z}}\right),
\end{align}
and the result is
\begin{align}
\notag \mathcal{H}_{T,2}\left(x,\xi,t,p^{z}\right)&=	\frac{\alpha_{s}C_F}{2\pi}\frac{1}{1-x}\begin{cases}
\frac{x+\xi}{1+\xi}\ln\frac{x-1}{x+\xi}+\frac{x-\xi}{1-\xi}\ln\frac{x-1}{x-\xi}+1 & x<-\xi\\
\frac{x+\xi}{1+\xi}\ln\frac{\left(p^{z}\right)^{2}}{m^{2}}+\frac{x-\xi}{1-\xi}\ln\frac{1-x}{\xi-x}+\frac{1-x}{1+\xi}\\
-\frac{x+\xi}{1+\xi}\ln\frac{1-x}{4\left(1+\xi\right)^{2}\left(x+\xi\right)} & -\xi<x<\xi\\
\frac{2\left(x-\xi^{2}\right)}{1-\xi^{2}}\left(\ln\frac{\left(p^{z}\right)^{2}}{m^{2}}-1\right)-\frac{x+\xi}{1+\xi}\ln\frac{1-x}{4\left(1+\xi\right)^{2}\left(x+\xi\right)}\\
-\frac{x-\xi}{1-\xi}\ln\frac{1-x}{4\left(1-\xi\right)^{2}\left(x-\xi\right)}+1 & \xi<x<1\\
-\frac{x+\xi}{1+\xi}\ln\frac{x-1}{x+\xi}-\frac{x-\xi}{1-\xi}\ln\frac{x-1}{x-\xi}-1 & x>1,
\end{cases}\\
\tilde{\mathcal{H}}_{T,2}\left(x,\xi,t,p^{z}\right)&=	0, \non\\
\mathcal{E}_{T,2}\left(x,\xi,t,p^{z}\right)&=	0, \non\\
\tilde{\mathcal{E}}_{T,2}\left(x,\xi,t,p^{z}\right)&=
\mathcal{O}\left(\frac{m^2}{\left(p^z\right)^2}\right).
\end{align}

The third term in the gluon propagator gives
\begin{align}
\notag \Gamma_{3}=&	C_{F}\int\frac{d^{4}k}{\left(2\pi\right)^{4}}\bar{u}\left(p+\frac{\Delta}{2}\right)\left(-ig_{s}\gamma^{\mu}\right)\frac{i}{k\!\!\!/+\frac{\Delta\!\!\!/}{2}-m}i\sigma^{z\perp}\frac{i}{k\!\!\!/-\frac{\Delta\!\!\!/}{2}-m}\left(-ig_{s}\gamma^{\nu}\right)u\left(p-\frac{\Delta}{2}\right)\\
	\notag &\times\frac{i\left[\left(p-k\right)^{\mu}\left(p-k\right)^{\nu}\right]}{\left[n\cdot\left(p-k\right)\right]^{2}\left(p-k\right)^{2}}\delta\left(x-\frac{k^{z}}{p^{z}}\right)\\
=&	C_{F}g_{s}^{2}\int\frac{d^{4}k}{\left(2\pi\right)^{4}}\bar{u}\left(p+\frac{\Delta}{2}\right)i\sigma^{z\perp}u\left(p-\frac{\Delta}{2}\right)\frac{1}{\left(p^{z}-k^{z}\right)^{2}\left(p-k\right)^{2}}\delta\left(x-\frac{k^{z}}{p^{z}}\right).
\end{align}
It contributes to $\mathcal{H}_T$ only with
\begin{align}
{\cal H}_{T,3}(x,\xi,t,\mu,p^{z})=	\frac{\alpha_{S}C_{F}}{2\pi}\frac{\sqrt{\mu^{2}+p_{z}^{2}\left(1-x\right)^{2}}-\left|1-x\right|p^{z}}{p^{z}\left(1-x\right)^{2}}.
\end{align}

Summing over all the above contributions, we obtain the complete result of the gluon exchange diagram in Fig. \ref{1loopGPD}
\begin{align}
\mathcal{H}_{T}^{\left(1\right)}\left(x,\xi,t,p^{z}\right)= & \frac{\alpha_{s}C_{F}}{2\pi}\begin{cases}
\ensuremath{\frac{1}{1-x}\left(\frac{x+\xi}{1+\xi}\ln\frac{x-1}{x+\xi}+\frac{x-\xi}{1-\xi}\ln\frac{x-1}{x-\xi}\right)+\frac{\mu}{(1-x)^{2}p^{z}}} & x<-\xi\\
\frac{\xi+x}{(1-x)(1+\xi)}\ln\frac{(p^{z})^{2}}{-t}\ensuremath{+\frac{1}{1-x}\left(\frac{\xi+x}{1+\xi}\ln\frac{\xi+x}{1-x}+\frac{\xi-x}{1-\xi}\ln\frac{\xi-x}{1-x}\right)}\\
-\frac{1}{(1-x)(1-\xi^{2})}\left(\left(\xi^{2}-x\right)\ln\frac{\xi-x}{\xi+x}+(1-x)\xi\ln\frac{(1+\xi)^{2}\left(\xi^{2}-x^{2}\right)}{4(1-x)^{2}\xi^{2}}\right)\\
+\frac{\xi+x}{(1-x)(1+\xi)}\left(\ln\left(4(1+\xi)^{2}\right)-1\right)+\frac{\mu}{(1-x)^{2}(p^{z})} & -\xi<x<\xi\\
\frac{2\left(x-\xi^{2}\right)}{(1-x)\left(1-\xi^{2}\right)}\ln\frac{\left(p^{z}\right)^{2}}{-t}\ensuremath{-\frac{2\left(x-\xi^{2}\right)}{(1-x)\left(1-\xi^{2}\right)}-\frac{2\xi}{1-\xi^{2}}\ln\frac{1+\xi}{1-\xi}}\\
-\frac{x+\xi}{(1-x)(1+\xi)}\ln\frac{1-x}{4(1+\xi)^{2}(\xi+x)}-\frac{x-\xi}{(1-x)(1-\xi)}\ln\frac{1-x}{4(1-\xi)^{2}(x-\xi)}\\
+\frac{\mu}{(-1+x)^{2}(p^{z})} & \xi<x<1\\
-\ensuremath{\frac{1}{1-x}\left(\frac{x+\xi}{1+\xi}\ln\frac{x-1}{x+\xi}+\frac{x-\xi}{1-\xi}\ln\frac{x-1}{x-\xi}\right)+\frac{\mu}{(1-x)^{2}p^{z}}} & x>1,
\end{cases}\non\\
\tilde{\mathcal{H}}_{T}\left(x,\xi,t,p^{z}\right)= & \mathcal{O}\left(\frac{m^{2}}{\left(p^{z}\right)^{2}}\right), \non\\
\mathcal{E}_{T}^{\left(1\right)}\left(x,\xi,t,p^{z}\right)= & \frac{\alpha_{s}C_{F}}{2\pi}\frac{2m^{2}}{-t}\begin{cases}
\frac{1}{1+\xi}\ln\frac{-t}{m^{2}}-\frac{2\xi}{1-\xi^{2}}\ln\frac{1+\xi}{2\xi(1-x)}\\
+\frac{1}{1+\xi}\ln(x+\xi)-\frac{1}{1-\xi}\ln(\xi-x) & -\xi<x<\xi\\
\frac{2}{1-\xi^{2}}\ln\frac{-t}{m^{2}}-\frac{2\xi}{1-\xi^{2}}\ln\frac{1+\xi}{1-\xi} & \xi<x<1\\
0 & \text{otherwise},
\end{cases}\non\\
\tilde{\mathcal{E}}_{T}^{\left(1\right)}\left(x,\xi,t,p^{z}\right)= & \frac{\alpha_{s}C_{F}}{2\pi}\frac{2m^{2}}{-t}\begin{cases}
\ensuremath{-\frac{1}{1+\xi}\ln\frac{-t}{m^{2}}-\frac{2}{1-\xi^{2}}\ln\frac{1+\xi}{2\xi(1-x)}\\
-\frac{1}{1+\xi}\ln(x+\xi)-\frac{1}{1-\xi}\ln(\xi-x)} & -\xi<x<\xi\\
\frac{2\xi}{1-\xi^{2}}\ln\frac{-t}{m^{2}}+\frac{2}{1-\xi^{2}}\ln\frac{1-\xi}{1+\xi} & \xi<x<1\\
0 & \text{otherwise}.
\end{cases}
\end{align}

The light cone quark transversity GPDs are obtained by taking the limit $p^z\to\infty$ before UV regularization. That is, we first integrate over $k^0$, then make a $p^z\rightarrow\infty$ expansion and integrate out $\vec{k}_\perp$.
The corresponding results for the three terms in the gluon propagator are
\begin{align}
\notag H_{T,1}\left(x,\xi,t\right)=	\frac{\alpha_{s}C_F}{2\pi}\begin{cases}
-\frac{x+\xi}{\left(1+\xi\right)\left(1-x\right)}\ln\frac{-t}{m^{2}}+\frac{x-\xi^{2}}{\left(1-x\right)\left(1-\xi^{2}\right)}\ln\frac{\xi-x}{\xi+x}\\
-\frac{\xi}{1-\xi^{2}}\ln\frac{\left(1+\xi\right)^{2}\left(\xi^{2}-x^{2}\right)}{4\left(1-x\right)^{2}\xi^{2}} & -\xi<x<\xi\\
-\frac{2\left(x-\xi^{2}\right)}{\left(1-\xi^{2}\right)\left(1-x\right)}\ln\frac{-t}{m^{2}}-\frac{2\xi}{1-\xi^{2}}\ln\frac{1+\xi}{1-\xi} & \xi<x<1\\
0 & \text{Otherwise},
\end{cases}
\end{align}

\begin{align}
H_{T,2}\left(x,\xi,t\right)&=	\frac{\alpha_{s}C_F}{2\pi}\begin{cases}
\frac{\left(x+\xi\right)}{\left(1+\xi\right)\left(1-x\right)}\left(\ln\frac{\mu^{2}}{m^{2}}+2\ln\frac{1+\xi}{1-x}\right) & -\xi<x<\xi\\
\frac{2\left(x-\xi^{2}\right)}{\left(1-\xi^{2}\right)\left(1-x\right)}\ln\frac{\mu^{2}}{m^{2}}+\frac{2\left(x+\xi\right)}{\left(1+\xi\right)\left(1-x\right)}\ln\frac{1+\xi}{1-x}\\
+\frac{2\left(x-\xi\right)}{\left(1-\xi\right)\left(1-x\right)}\ln\frac{1-\xi}{1-x} & \xi<x<1\\
0 & \text{Otherwise},
\end{cases}\non\\
\tilde{H}_{T,\{1,2\}}\left(x,\xi,t\right)&= \mathcal{O}\left(\frac{m^2}{\left(p^z\right)^2}\right).
\end{align}
$E_T$ and $\tilde{E}_T$ are the same as $\mathcal{E}_T$ and $\tilde{\mathcal{E}}_T$, respectively. The third term in the gluon propagator clearly vanishes on the light cone.
Summing over all these contributions leads to the following light cone result
\begin{align}
H_{T}^{\left(1\right)}\left(x,\xi,t,p^{z}\right)= & \frac{\alpha_{s}C_{F}}{2\pi}\begin{cases}
\frac{x+\xi}{\left(1+\xi\right)\left(1-x\right)}\ln\frac{\mu^{2}}{-t}-\frac{\xi}{1-\xi^{2}}\ln\frac{(\xi+1)^{2}\left(\xi^{2}-x^{2}\right)}{4\xi^{2}(1-x)^{2}}\\
+\frac{x-\xi^{2}}{\left(1-\xi^{2}\right)\left(1-x\right)}\ln\frac{\xi-x}{\xi+x}+\frac{2\left(x+\xi\right)}{(1+\xi)(1-x)}\ln\frac{1+\xi}{1-x} & -\xi<x<\xi\\
\frac{2\left(x-\xi^{2}\right)}{\left(1-\xi^{2}\right)(1-x)}\ln\frac{\mu^{2}}{-t}-\frac{2\xi}{1-\xi^{2}}\ln\frac{1+\xi}{1-\xi}\\
+\frac{2(x-\xi)}{(1-\xi)(1-x)}\ln\frac{1-\xi}{1-x}+\frac{2(x+\xi)}{(\xi+1)(1-x)}\ln\frac{1+\xi}{1-x} & \xi<x<1\\
0 & \text{otherwise},
\end{cases}\non\\
\tilde{H}_{T}\left(x,\xi,t,p^{z}\right)= & \mathcal{O}\left(\frac{m^{2}}{\left(p^{z}\right)^{2}}\right),\non\\
E_{T}^{\left(1\right)}\left(x,\xi,t,p^{z}\right)= & \tilde{\mathcal{E}}_{T}^{\left(1\right)}\left(x,\xi,t,p^{z}\right),\non\\
\tilde{E}^{\left(1\right)}_{T}\left(x,\xi,t,p^{z}\right)= & \tilde{\mathcal{E}}^{\left(1\right)}_{T}\left(x,\xi,t,p^{z}\right).
\end{align}
As in the case of unpolarized and longitudinally polarized quark GPDs, the quasi result does not vanish in the full $x$ region, whereas the light cone result is nonvanishing only in the Efremov-Radyushkin-Brodsky-Lepage (ERBL) and DokshitzerGribovLipatovAltarelliParisi (DGLAP) region $-\xi<x<\xi$ and $\xi<x<1$. Moreover, there is no $\ln\mu$ dependence in the quasi result, instead it contains a $\ln p^z$ dependence. The reason is that the limit $p^z\to\infty$ does not commute with the UV regularization. However, the $\ln p^z$ term in the quasi distribution has the same coefficient as the $\ln\mu$ term in the light cone distribution, the former can actually be traded into the latter by a matching factor of the form $\ln{p^z/\mu}$.

The quark self-energy diagrams in Fig.~\ref{1loopGPD} lead to the quark wave function renormalization factor, which has been calculated in Ref.~\cite{Ji:2015qla}. For completeness, we list the result here
\begin{align}
{\cal Z}_{F}^{(1)} & =-\frac{\alpha_{S}C_{F}}{2\pi}\int dy\begin{cases}
\big(f(\xi,y)\ln\frac{y-\xi}{y-1}+f(-\xi,y)\ln\frac{y+\xi}{y-1}\big)-\frac{1}{1-\xi^{2}}+\frac{\mu}{p^{z}(1-y)^{2}} & y<-\xi\\
-f(-\xi,y)\ln\frac{p_{z}^{2}}{m^{2}}-f(\xi,y)\ln\frac{1-y}{\xi-y}+f(-\xi,y)\ln\frac{1-y}{4(1+\xi)^{2}(\xi+y)}\\
+4f(-\xi,y)-\frac{1}{1-\xi^{2}}+\frac{2}{1-y}+\frac{\mu}{p^{z}(1-y)^{2}} & -\xi<y<\xi\\
-(f(\xi,y)+f(-\xi,y))\ln\frac{p_{z}^{2}}{m^{2}}+f(\xi,y)\ln\frac{1-y}{4(y-\xi)(1-\xi)^{2}}\\
+f(-\xi,y)\ln\frac{1-y}{4(y+\xi)(1+\xi)^{2}}+4(f(\xi,y)+f(-\xi,y))+\frac{4}{1-y}-\frac{1}{1-\xi^{2}}\\
+\frac{\mu}{p^{z}(1-y)^{2}} & \xi<y<1\\
-f(\xi,y)\ln\frac{y-\xi}{y-1}-f(-\xi,y)\ln\frac{y+\xi}{y-1}+\frac{1}{1-\xi^{2}}+\frac{\mu}{p^{z}(1-y)^{2}} & y>1,
\end{cases}\non
\end{align}
\begin{align}
Z_{F}^{(1)} & =-\frac{\alpha_{S}C_{F}}{2\pi}\int dy\begin{cases}
-f(-\xi,y)\ln\frac{\mu^{2}}{m^{2}}-2f(-\xi,y)\ln\frac{1+\xi}{1-y}+\frac{1}{1+\xi}-\frac{1}{1-y} & -\xi<y<\xi\\
-(f(\xi,y)+f(-\xi,y))\ln\frac{\mu^{2}}{m^{2}}-2f(\xi,y)\ln\frac{1-\xi}{1-y}\\
-2f(-\xi,y)\ln\frac{1+\xi}{1-y}+\frac{1}{1-\xi}+\frac{1}{1+\xi}-\frac{2}{1-y} & \xi<y<1\\
0 & \text{otherwise},
\end{cases}
\end{align}
where ${\cal Z}_{F}^{(1)}, Z_{F}^{(1)}$ are for the quasi and the light cone distribution, respectively, and
\begin{align}
f(\xi,y) & =\frac{1}{1-\xi}-\frac{1}{1-y}-\frac{1-y}{2(1-\xi)^{2}}.
\end{align}



\section{one-loop factorization}
Following Refs.~\cite{Xiong:2013bka, Ji:2015qla}, the factorization connecting the quasi GPD $\mathcal{H}_{T}$ and the light cone
GPD $H_{T}$ can be written as
\begin{align}
\mathcal{H}_{T}\left(x,\xi,t,\mu,p^{z}\right)= & \int_{-1}^{1}\frac{dy}{\left|y\right|}Z_{H_{T}}\left(\frac{x}{y},\frac{\xi}{y},\frac{\mu}{p^{z}}\right)H_{T}\left(y,\xi,t,\mu\right)
\end{align}
up to power corrections suppressed by $p^z$, where the integration range is given by the support property of the light cone GPD. The matching factor $Z_{H_{T}}$ is completely perturbative, and can be expanded as
\begin{align}
Z_{H_{T}}\left(\frac{x}{y},\frac{\xi}{y},\frac{\mu}{p^{z}}\right)= & \delta\left(1-\frac{x}{y}\right)+\frac{\alpha_{s}}{2\pi}Z_{H_{T}}^{\left(1\right)}\left(\frac{x}{y},\frac{\xi}{y},\frac{\mu}{p^{z}}\right)+h.o.
\end{align}
where h.o. denotes higher-order corrections. Substituting the
above expansion of $Z_{H_T}$ into the matching condition, one obtains
\begin{align}
Z_{H_{T}}^{\left(1\right)}\left(\eta,\zeta,\mu,p^{z}\right)= & \frac{2\pi}{\alpha_{s}}\left[\mathcal{H}_{T}^{(1)}\left(\eta,\zeta,p^{z}\right)-H_{T}^{(1)}\left(\eta,\zeta,\mu\right)\right],
\end{align}
which is given by
\begin{align}
Z_{H_{T}}^{\left(1\right)}\left(\eta,\zeta,\mu/p^{z}\right)/C_{F}= & \begin{cases}
\frac{1}{1-\eta}\left(\frac{\eta+\zeta}{1+\zeta}\ln\frac{\eta-1}{\eta+\zeta}+\frac{\eta-\zeta}{1-\zeta}\ln\frac{\eta-1}{\eta-\zeta}\right)+\frac{\mu}{(1-\eta)^{2}p^{z}} & \eta<-\zeta\\
\ensuremath{\frac{\eta+\zeta}{(1+\zeta)(1-\eta)}\ln\frac{\left(p^{z}\right)^{2}}{\mu^{2}}+\frac{1}{1-\eta}\left(\frac{\eta+\zeta}{1+\zeta}\ln\frac{\eta+\zeta}{1-\eta}+\frac{\zeta-\eta}{1-\zeta}\ln\frac{\zeta-\eta}{1-\eta}\right)}\\
+\frac{\eta+\zeta}{\left(1-\eta\right)\left(1+\zeta\right)}\left(\ln(4(1-\eta)^{2})-1\right)+\frac{\mu}{(1-\eta)^{2}p^{z}} & -\zeta<\eta<\zeta\\
\frac{2\left(\eta-\zeta^{2}\right)}{(1-\eta)\left(1-\zeta^{2}\right)}\left(\ln\frac{\left(p^{z}\right)^{2}}{\mu^{2}}+\ln\left(4\left(1-\eta\right)^{2}\right)-1\right)\\
+\frac{\eta+\zeta}{\left(1-\eta\right)\left(1+\zeta\right)}\ln\frac{\eta+\zeta}{1-\eta}+\frac{\eta-\zeta}{\left(1-\eta\right)\left(1-\zeta\right)}\ln\frac{\eta-\zeta}{1-\eta}+\frac{\mu}{(1-\eta)^{2}p^{z}} & \zeta<\eta<1\\
-\ensuremath{\frac{1}{1-\eta}\left(\frac{\eta+\zeta}{1+\zeta}\ln\frac{\eta-1}{\eta+\zeta}+\frac{\eta-\zeta}{1-\zeta}\ln\frac{\eta-1}{\eta-\zeta}\right)+\frac{\mu}{(1-\eta)^{2}p^{z}}} & \eta>1.
\end{cases}\label{eq:HT_matching}
\end{align}
The above result is valid only for $y>\xi$. However, it
can be extended to the whole $y$ region as
\begin{align}
\frac{1}{|y|}Z_{H_{T}}^{(1)}\left(\frac{x}{y},\frac{\xi}{y},\frac{\mu}{p^{z}}\right)/C_{F} & =\frac{1}{y}\Big[Z_{H_{T}, 1}^{(1)}\left(\frac{x}{y},\frac{\xi}{y},\frac{\mu}{p^{z}}\right)\theta(x<-\xi)\theta(x<y)\nonumber \\
 & +Z_{H_{T}, 2}^{(1)}\left(\frac{x}{y},\frac{\xi}{y},\frac{\mu}{p^{z}}\right)\theta(-\xi<x<\xi)\theta(x<y)\nonumber \\
 & +Z_{H_{T}, 3}^{(1)}\left(\frac{x}{y},\frac{\xi}{y},\frac{\mu}{p^{z}}\right)\theta(\xi<x<y)+Z_{H_{T}, 4}^{(1)}\left(\frac{x}{y},\frac{\xi}{y},\frac{\mu}{p^{z}}\right)\theta(x>\xi)\theta(x>y)\Big],\label{eq:All_y_match_fac}
\end{align}
where $Z_{H_{T}, i}^{(1)}$ are the analytic continuation of the above matching
factor in the four different regions in Eq.\ (\ref{eq:HT_matching}) so
that they are real functions (one simply needs to replace $\ln a$ with $1/2\ln a^2$) and $p^{z}$ shall be replaced by $yP^{z}$ with $P^{z}$ the averaged longitudinal momentum of the external hadrons. The validity of the above equation can be checked by explicit computations.


The wave function renormalization factor introduces an extra contribution
to the matching factor near $\eta=x/y=1$:
\begin{align}
Z_{H_{T}}^{(1)}\left(\frac{x}{y},\frac{\xi}{y},\frac{\mu}{p^{z}}\right)= & \delta\left(1-\eta\right)\delta Z_{H}^{(1)}/\left(\frac{\alpha_{s}}{2\pi}\right),
\end{align}
where $\delta Z_{H}^{(1)}$ has been calculated in Ref.~\cite{Ji:2015qla}:
\begin{align}
\delta Z_{H}^{(1)} & =-\frac{\alpha_{S}C_{F}}{2\pi}\int d\eta\begin{cases}
\big(f(\zeta,\eta)\ln\frac{\eta-\zeta}{\eta-1}+f(-\zeta,\eta)\ln\frac{\eta+\zeta}{\eta-1}\big)-\frac{1}{1-\zeta^{2}}+\frac{\mu}{p^{z}(1-\eta)^{2}} & \eta<-\zeta\\
-f(-\zeta,\eta)\ln\frac{p_{z}^{2}}{\mu^{2}}-f(\zeta,\eta)\ln\frac{1-\eta}{\zeta-\eta}-f(-\zeta,\eta)\ln[4(\zeta+\eta)(1-\eta)]\\
+4f(-\zeta,\eta)-\frac{1}{1-\zeta^{2}}-\frac{1}{1+\zeta}+\frac{3}{1-\eta}+\frac{\mu}{p^{z}(1-\eta)^{2}} & -\zeta<\eta<\zeta\\
-(f(\zeta,\eta)+f(-\zeta,\eta))\ln\frac{p_{z}^{2}}{\mu^{2}}-f(\zeta,\eta)\ln[4(\eta-\zeta)(1-\eta)]\\
-f(-\zeta,\eta)\ln[4(\zeta+\eta)(1-\eta)]+4(f(\zeta,\eta)+f(-\zeta,\eta))+\frac{6}{1-\eta}\\
-\frac{3}{1-\zeta^{2}}+\frac{\mu}{p^{z}(1-\eta)^{2}} & \zeta<\eta<1\\
-f(\zeta,\eta)\ln\frac{\eta-\zeta}{\eta-1}-f(-\zeta,\eta)\ln\frac{\eta+\zeta}{\eta-1}+\frac{1}{1-\zeta^{2}}+\frac{\mu}{p^{z}(1-\eta)^{2}} & \eta>1.
\end{cases}\label{1loopmatfacwf}
\end{align}
Now we have the complete one-loop matching factor for $H_T$. As for $E_T$ and $\tilde E_T$, we can see from the results in the previous section that the matching factor is a trivial $\delta$ function
\begin{align}
Z_{E_{T}}^{(1)}\left(\eta,\zeta\right)=Z_{\tilde{E}_{T}}^{(1)}\left(\eta,\zeta\right)= & \delta\left(\eta-1\right)
\end{align}
up to one-loop level and leading $p^z$ accuracy. The reason is that the light cone GPDs $E_{T}$, ${\tilde{E}}_{T}$ vanish at tree level, and therefore must be UV finite at one-loop level. They do not have a cutoff dependence. Accordingly, $\mathcal{E}_{T}$ and $\mathcal{\tilde{E}}_{T}$ do not have a logarithmic dependence on $p^z$. $E_{T}$ and ${\tilde{E}}_{T}$ can therefore be smoothly approached by the large momentum limit of their quasi counterparts.

\section{conclusion}

We have presented the one-loop matching condition for the quark transversity GPD in the nonsinglet case. The matching factor for the GPD $H_T$ is nontrivial, and reduces to that for the transversity quark distribution in the forward limit. The matching factor for $E_T$ and $\tilde E_T$ is a trivial $\delta$ function to one-loop and leading power accuracy. Both ${\tilde H}_T$ and its quasi counterpart $\tilde{\mathcal{H}}_{T}$ are power suppressed by the hadron momentum, and therefore are omitted at leading power accuracy.

\section{Acknowledgments}
We thank Barbara Pasquini and Peter Schweitzer for the discussion on evolution of chiral-odd GPDs. We also thank V. Braun and A. Sch\"afer for useful discussions. This work was partially supported by a DFG grant SCHA 458/20-1 and a grant from National Science Foundation of China (No.
11405104).

\end{document}